\documentclass[aps,showpacs,amsfonts,nofootinbib,preprintnumbers,superscriptaddress,nobalancelastpage]{revtex4-2}
\pdfoutput=1
\usepackage{graphicx}
\usepackage{amsmath,amssymb}
\usepackage{url}
\usepackage{multirow}
\usepackage{hyperref,color}
\usepackage[normalem]{ulem}
\usepackage{slashed}
\usepackage{rotating}
\usepackage{afterpage}
\usepackage{ifthen}
\usepackage{mciteplus}
\usepackage{multirow}
\usepackage{color}

\def\lsim{\raise0.3ex\hbox{$\;<$\kern-0.75em\raise-1.1ex\hbox{$\sim\;$}}}
\def\gsim{\raise0.3ex\hbox{$\;>$\kern-0.75em\raise-1.1ex\hbox{$\sim\;$}}}

\newcommand{\UH}{\mathbf{U}}

\newcommand{\DLR}{\mathbf{D}}
\def\Tr{{\rm Tr}}

\begin{document}

\preprint{YITP-SB-2023-38}

\title{Bounds on Quartic Gauge Couplings in HEFT from
 Electroweak Gauge Boson Pair Production at the LHC}
\author{O.\ J.\ P.\ \'Eboli}
\email{eboli@if.usp.br}
\affiliation{Instituto de F\'{\i}sica, 
Universidade de S\~ao Paulo, S\~ao Paulo -- SP, Brazil.}
\author{M.~C.~Gonzalez-Garcia,}
\email{maria.gonzalez-garcia@stonybrook.edu}
\affiliation{Departament de Fis\'{\i}ca Qu\`antica i
  Astrof\'{\i}sica and Institut de Ciencies del Cosmos, Universitat de
  Barcelona, Diagonal 647, E-08028 Barcelona, Spain}
\affiliation{Instituci\'o Catalana de Recerca i Estudis
  Avan\c{c}ats (ICREA), Pg. Lluis Companys 23, 08010 Barcelona,
  Spain.}
\affiliation{C.N. Yang Institute for Theoretical Physics,
  Stony Brook University, Stony Brook New York 11794-3849, USA}
\author{Matheus Martines}
\email{matheus.martines.silva@usp.br}
\affiliation{Instituto de F\'{\i}sica, 
Universidade de S\~ao Paulo, S\~ao Paulo -- SP, Brazil.}

\begin{abstract}

  Precision measurements of anomalous quartic couplings of electroweak
  gauge bosons allow us to search for deviations of the Standard Model
  predictions and signals of new physics. Here, we obtain the
  constraints on anomalous quartic gauge couplings using the presently
  available data on the production of gauge-boson pairs via vector
  boson fusion. We work in the Higgs effective theory framework and
  obtain the present bounds on the operator's Wilson
  coefficients. Anomalous quartic gauge boson couplings lead to
  rapidly growing cross sections and we discuss the impact of a
  unitarization procedure on the attainable limits.

\end{abstract}

\maketitle

\section{Introduction}

The Standard Model (SM) $SU(2)_L \otimes U(1)_Y$ gauge symmetry
determines univocally the structure and strength of the triple and
quartic couplings among electroweak gauge-bosons.  Therefore,
measuring independently the triple gauge-boson couplings (TGC) and the
quartic gauge-boson couplings (QGC) tests the SM and provides
sensitivity to new physics. In a model independent approach,
departures from the SM predictions for TGC and QGC can be parametrized
by higher-order operators encoding indirect effects of heavy new
physics. Furthermore, the analysis of the gauge boson self
interactions can probe whether the gauge symmetry is realized linearly
or nonlinearly in the low energy effective theory (EFT) of the
electroweak symmetry breaking sector~\cite{Brivio:2016fzo,
  Brivio:2014pfa}.  \smallskip

In collider experiments, the pair production of electroweak gauge
bosons allows the direct study of TGC~\cite{Brown:1978mq,
  Hagiwara:1986vm, Baur:1989gk}, while QGC can be probed via the
production of three electroweak vector bosons~\cite{Belanger:1992qh,
  Dervan:1999as, Eboli:2000ad, Aad:2015uqa, Chatrchyan:2014bza,
  Aaboud:2017tcq, Sirunyan:2017lvq}, the exclusive production of
gauge-boson pairs ~\cite{ Belanger:1992qi, Chatrchyan:2013akv,
  Khachatryan:2016mud}, or the vector-boson-scattering production of
electroweak vector boson pairs~\cite{ Belyaev:1998ih,
  Eboli:2000ad,Eboli:2003nq, Khachatryan:2016vif, Aaboud:2016ffv,
  Khachatryan:2017jub, Sirunyan:2017fvv, Sirunyan:2019der,
  Sirunyan:2020tlu,CMS:2020fqz,CMS:2020gfh, Hwang:2023wad}.  In the
EFT approach, the Wilson coefficients of effective operators that
contain both TGC and QGC are more strongly constrained through the
study of their TGC component.  \smallskip

In order to mitigate the bounds on QGC originating from the TGC
analyses, we focus on the so-called {\sl genuine} QGC operators, that
is, effective operators generating QGC but that do not generate any
TGC; for models leading to such operators see~\cite{Godfrey:1995cd},
for instance.  The set of operators to be considered depends on the
assumed realization of the SM gauge theory in the low-energy EFT in
which the nature of the Higgs-like state observed at the LHC in
2012~\cite{Aad:2012tfa, Chatrchyan:2012xdj} plays a pivotal role. If
the Higgs belongs to a $SU(2)_L$ doublet, the SM gauge symmetry can be
realized linearly in the effective theory, which, in this case, is
usually referred to as {\sl standard model effective field theory}
(SMEFT).  In this scenario, the lowest-order genuine QGC are given by
dimension-eight operators~\cite{Eboli:2006wa}.  Alternatively, if the
Higgs boson is a $SU(2)_L$ isosinglet, we are lead to use a nonlinear
realization of the gauge symmetry and the low energy EFT obtained this
way is called {\sl Higgs effective theory} (HEFT).  In this case, the
lowest-order QGC appear at
${\cal O}(p^4)$~\cite{Alonso:2012px,Brivio:2013pma}. \smallskip

There is one important difference between the QGC generated 
gauge-linear dimension-eight operators and those generated nonlinearly
at ${\cal O}(p^4)$: in the second case the QGC's do not
involve photons. This fact renders these operators more difficult to
observe, specially in the production of three gauge bosons.
Consequently, most of the experimental searches have casted their
results on QGC as bounds on Wilson coefficients of dimension-eight
gauge-linear operators.  Furthermore, most experimental searches
consider only one Wilson coefficient different from zero at a
time. This implies that the results of the experimental searches
constraining dimension-eight SMEFT operators, even those which do not
involve photons nor derivatives, cannot be directly translated into
bounds on the ${\cal O}(p^4)$ HEFT operators because the last ones are
equivalent to combinations of several coefficients of the
corresponding dimension-eight SMEFT siblings; see next section for
details. \smallskip

With this motivation, in this work we perform a dedicated combined
analysis of searches for genuine QGC in the framework of the
${\cal O}(p^4)$ HEFT operators. We briefly present in
Sec.~\ref{sec:frame} the basics of the analysis framework.  We focus
on the most sensitive channels for the generated QGC which are those
with electroweak gauge boson pairs produced in association with two
jets, which are dominated by vector boson fusion.
Section.~\ref{sec:ana} describes the data sets considered and the
details of our analysis, while we present our results and their
discussion in Section~\ref{sec:res}.  \smallskip

\section{Analysis Framework}
\label{sec:frame}

In this work we consider a dynamical scenario in which the Higgs boson
is a pseudo-Nambu-Goldstone boson of a broken global symmetry while
being an isosinglet of the SM gauge symmetries.  In this case, the
gauge symmetry of the low energy effective Lagrangian is realized
nonlinearly with a global $SU(2)_L \otimes SU(2)_R$ symmetry broken to
the diagonal $SU(2)_C$~\cite{Weinberg:1978kz, Feruglio:1992wf,
  Appelquist:1980vg, Longhitano:1980tm}. This EFT is a derivative
expansion and it is written in terms of the SM fermions and gauge
bosons and of the physical Higgs
$h$~\cite{Brivio:2016fzo,Alonso:2012px}.  The building block at low
energies is a dimensionless unitary matrix transforming as a
bi-doublet of the global symmetry $SU(2)_L \otimes SU(2)_R$:
\begin{equation}
\UH(x)=e^{i\sigma_a \pi^a(x)/v}\qquad\qquad , \qquad \qquad  \UH(x) \rightarrow L\, \UH(x) R^\dagger\;,
\end{equation}
where $L$, $R$ denote $SU(2)_{L,R}$ global transformations,
respectively and $\pi^a$ are the Goldstone bosons.  Its covariant
derivative is given by
\begin{equation}
\DLR_\mu \UH(x) \equiv \partial_\mu \UH(x) +ig \frac{\sigma^j}{2}
W^i_{\mu}(x)\UH(x) - \frac{ig'}{2}  B_\mu(x) \UH(x)\sigma_3 \; .
\end{equation}
From this basic element it is possible to construct the vector chiral
field $V_\mu$ and the scalar chiral field $T$ that transform in the
adjoint of $SU(2)_L$
\begin{equation}
  V_\mu \equiv   \left(\DLR_\mu\UH\right)\UH^\dagger \;\;\;,\;\;\;
  T\equiv\UH\sigma_3\UH^\dag \;.
\end{equation}

The lowest order genuine quartic operators are ${\cal O}(p^4)$ which
require only two building blocks~\cite{Eboli:2016kko}
\begin{equation}
  {\rm Tr} \left [ T V_\mu \right ] = i \frac{g}{c_W} Z_\mu
  \;\;\;\hbox{ and }\;\;\;
  {\rm Tr}\left [ V_\mu V_\nu \right ] = - \frac{g^2}{2} \left (
    \frac{1}{c^2_W} Z_\mu Z_\nu + W^+_\mu W^-_ nu + W^+_\nu W^-_\mu
  \right ) \;.
\end{equation}
At this order, there are two operators which respect the $SU(2)_C$
custodial symmetry, as well as $C$ and $P$, that in the notation of
Refs.~\cite{Alonso:2012px,Brivio:2013pma}, are
\begin{equation}
\begin{array}{l}
{\cal P}_6 = \Tr [ V^\mu V_\mu] \Tr[V^\nu V_\nu] {\cal F}_6(h)
= g^4\left[\frac{1}{4c_w^4}  {\cal O}^{\partial= 0}_{ZZ}
+ {\cal O}^0_{WW,1} +\frac{1}{c_w^2}  {\cal O}^{\partial=0}_{WZ,1}\right] {\cal F}_6(h)\; , 
\\
{\cal P}_{11} = \Tr [ V^\mu V^\nu] \Tr[V_\mu V_\nu] {\cal F}_{11}(h)
=g^4\left[
\frac{1}{4c_w^4}  {\cal O}^{\partial =0}_{ZZ}
+ \frac{1}{2} {\cal O}^{\partial=0}_{WW,1}
+ \frac{1}{2} {\cal O}^{\partial=0}_{WW,2}
+ \frac{1}{c_w^2}  {\cal O}^{\partial =0}_{WZ,2}\right]{\cal F}_{11}(h) \;,  
\label{eq:p41}
\end{array}
\end{equation}
and 3 additional $CP$ conserving operators that violate $SU(2)_C$:
\begin{equation}
\begin{array}{l}
{\cal P}_{23} = \Tr [ V^\mu V_\mu] (\Tr[ T V_\nu])^2 {\cal F}_{23}(h)
=g^4\left[\frac{1}{2c_w^4}  {\cal O}^{\partial=0}_{ZZ}
+ \frac{1}{c_w^2}  {\cal O}^{\partial=0}_{WZ,1}\right]{\cal F}_{23}(h) \; , 
\\
{\cal P}_{24} = \Tr [ V^\mu V^\nu] \Tr[ T V_\mu] \Tr[T V_\nu]
{\cal F}_{24}(h)
=g^4\left[\frac{1}{2c_w^4}  {\cal O}^{\partial=0}_{ZZ}
+ \frac{1}{c_w^2}  {\cal O}^{\partial=0}_{WZ,2}\right]{\cal F}_{24}(h)\; , 
\\ 
{\cal P}_{26} = ( \Tr[ T V_\mu] \Tr[T V_\nu] )^2{\cal F}_{26}(h)
= \frac{g^4}{c_w^4}  {\cal O}^{\partial=0}_{ZZ}{\cal F}_{26}(h)
\; ,
\label{eq:p42}
\end{array}
\end{equation}
which we have expressed in terms five  basic four gauge-boson vertices
\begin{equation}
\begin{array}{lcl}
 {\cal Q}^{\partial=0}_{\text WW,1} = W^{+ \mu} W^-_\mu W^{+ \nu} W^-_\nu 
&,
& {\cal Q}^{\partial=0}_{\text WW,2} = W^{+ \mu} W^{-\nu} W^+_\mu W^-_\nu  \;,
\\
 {\cal Q}^{\partial=0}_{\text WZ,1} = W^{+ \mu} W^-_\mu Z^{\nu} Z_\nu 
&,
& {\cal Q}^{\partial=0}_{\text WZ,2} = W^{+ \mu} W^{-\nu} Z_\mu Z_\nu \;,
\\
{\cal Q}^{\partial=0}_{\text ZZ} = Z^\mu Z_\mu Z^\nu Z_\nu \;.
&
\end{array}
\label{eq:0deriv}
\end{equation}
In addition, ${\cal F}_i(h)$ are generic functions parametrizing the
chiral-symmetry breaking interactions of $h$. As we are looking for
operators whose lowest order vertex contain four gauge bosons, we take
${\cal F}_i=1$. \smallskip

As mentioned in the introduction, the above operators do not contain
photons. We also see that there are five operators matching five
independent Lorentz structures that do not exhibit derivatives.  These
two facts make these operators more difficult to bound.  The first
four structures in Eq.~(\ref{eq:0deriv}) modify the SM quartic
couplings $W^+W^-W^+W^-$ and $W^+W^-ZZ$, while the last one leads to
$ZZZZ$ QGC not present in the SM. \smallskip

The most general effective Lagrangian at ${\cal O}(p^4)$ for genuine
QGC is
\begin{equation}
{\cal L}^{p=4}_{QGC}=\sum_{i=6,11,23,24,26} \,C_i\, {\cal P}_i \;.
\end{equation} 
In Ref.~\cite{Longhitano:1980tm} we can also find the ${\cal O}(p^4)$
QGC assuming that there is no light Higgs-like state and this
corresponds to the limit ${\cal F}_i \to 1$ in our framework.  The
translation between the Wilson coefficients our notation and the one
of Ref.~\cite{Longhitano:1980tm} is
\begin{equation}
\alpha_4 = C_{11} \;\;,\;\;
\alpha_5 = C_{6} \;\;,\;\;
\alpha_6 = C_{24} \;\;,\;\;
\alpha_7 = C_{23}\;\;,\;\;
\alpha_{10} =C_{26}\;\;.
\end{equation}

Let us finish by listing the corresponding sub-set of dimension-8
operators of the SMEFT which do not involve derivatives of gauge
fields. There are three of those
\begin{equation}
\begin{array}{l}
  {\cal O}_{S,0} = 
\left [ \left ( D_\mu \Phi \right)^\dagger
 D_\nu \Phi \right ] \times 
\left [ \left ( D^\mu \Phi \right)^\dagger
D^\nu \Phi \right ]
=\frac{g^{{4}} v^4}{16} \left[
 {\cal Q}^{\partial=0}_{WW,2}
+\frac{1}{c_w^2}  {\cal Q}^{\partial=0}_{WZ,2} 
+\frac{1}{4c_w^4}  {\cal Q}^{\partial=0}_{ZZ}  \right]
\; , 
\\
  {\cal O}_{S,1} =
 \left [ \left ( D_\mu \Phi \right)^\dagger
 D^\mu \Phi  \right ] \times
\left [ \left ( D_\nu \Phi \right)^\dagger
D^\nu \Phi \right ]
= \frac{g^4 v^{{4}}}{16} \left[
 {\cal Q}^{\partial=0}_{WW,1}
+\frac{1}{c_w^2}  {\cal Q}^{\partial=0}_{WZ,1} 
+\frac{1}{4c_w^4}  {\cal Q}^{\partial=0}_{ZZ}  \right]
\; , 
\\
  {\cal O}_{S,2} =
 \left [ \left ( D_\mu \Phi \right)^\dagger
 D_\nu \Phi  \right ] \times
\left [ \left ( D^\nu \Phi \right)^\dagger
D^\mu \Phi \right ]
= \frac{g^{4} v^4}{16} \left[
 {\cal Q}^{\partial=0}_{WW,1}
+\frac{1}{c_w^2}  {\cal Q}^{\partial=0}_{WZ,2} 
+\frac{1}{4c_w^4}  {\cal Q}^{\partial=0}_{ZZ}  \right]
\; . 
\end{array}
\label{eq:dphi}
\end{equation}
From the expressions above it is clear that, in general, the
constraints derived on the coefficients of these three operators
cannot be directly translated on bounds of the $C_i$ coefficients and
that a dedicated analysis is required, which we present next.
\smallskip

\section{Analysis of Electroweak Diboson Production in Association
  with Jets}
\label{sec:ana}

The electroweak production of $WZ$, $WW$ and $ZZ$ pairs in association
with two jets allow us to study the quartic couplings of electroweak
gauge bosons which contribute to the above processes via vector boson
fusion (VBF) .  In this work we consider the latest results on VBF
from CMS and ATLAS summarized in Table~\ref{tab:vbf-data} which
comprise a total of 18 data points.  For convenience, we also identify
in the table which operators contribute to each channel.  \smallskip

\begin{table} [ht]
\begin{tabular}{|l|l|c|l|c||ccccc|}
\hline
Channel ($a$) &Data set & Int Lum
& Distribution & \# bins
&  $\mathcal{P}_6$ &  $\mathcal{P}_{11}$ &  $\mathcal{P}_{23}$ &
$\mathcal{P}_{24}$
&  $\mathcal{P}_{26}$ 
\\ [0mm]
  \hline
  $ZZjj \to \ell^+ \ell^- \ell^{\prime +}\ell^{\prime -} jj$
 & CMS 13  TeV ~\cite{CMS:2020fqz}
  &   137 fb$^{-1}$  & $M_{4l}$ (Fig.4)  & 6
  &\checkmark & \checkmark & \checkmark &\checkmark & \checkmark 
  \\[0mm]
  $W^\pm W^\pm jj\to \ell^\pm \nu \ell^{\prime\pm}\nu j j$
  & CMS 13 TeV ~\cite{CMS:2020gfh}
  & 137 fb$^{-1}$
  &$M_T^{WW}$
  (Fig.6)
  &5
  &\checkmark & \checkmark & $\times$ &$\times$ & $\times$ 
  \\[0mm]
  $WZ jj
  \to \ell^+ \ell^- \ell^{\prime}\nu j j$
& CMS 13 TeV ~\cite{CMS:2020gfh}
  & 137 fb$^{-1}$
  &$M_T^{WZ}$  (Fig.6)
  &3 
&\checkmark & \checkmark & \checkmark &\checkmark & $\times$
  \\[0mm]
$ZZjj \to \ell^+ \ell^- \ell^{\prime +}\ell^{\prime -} jj$
& ATLAS 13  TeV ~\cite{ATLAS:2023dkz}
  &   140 fb$^{-1}$   & $d\sigma/dm_{4\ell}$
  (Fig.4)  & 4
  &\checkmark & \checkmark & \checkmark &\checkmark & \checkmark 
  \\[0mm]
  \hline
\end{tabular}
\caption{Data from LHC used in the analyses. In each case we list the
  figure of the distribution used in the analyses.  For the $ZZ jj$
  channel from CMS~\cite{CMS:2020fqz} we have merged the contents of
  the last three bins to ensure gaussianity.  For the $WZ jj$ channel
  from CMS~\cite{CMS:2020gfh} we only use the most sensitive bins that
  are the three highest invariant mass ones satisfying $M_T^{WZ}>700$
  GeV. For the $ZZjj$ channel in ATLAS ~\cite{ATLAS:2023dkz} following
  the collaboration we remove from the analysis the first of the 5
  bins of the $d\sigma/dm_{4l}$ distribution.}
\label{tab:vbf-data}
\end{table}

The theoretical prediction corresponding to the different data sets
are obtained by simulating at the required order $W^\pm W^\pm jj$,
$W^\pm Zjj$, $ZZjj$ events.  To this end, we use
\textsc{MadGraph5\_aMC@NLO}~\cite{Frederix:2018nkq} with the UFO files
for our effective Lagrangian generated with
\textsc{FeynRules}~\cite{Christensen:2008py, Alloul:2013bka}.  We
employ \textsc{PYTHIA8}~\cite{Sjostrand:2007gs} to decay the gauge
bosons and to perform the parton shower and hadronization, while the
fast detector simulation is carried out with
\textsc{Delphes}~\cite{deFavereau:2013fsa}.  Jet analyses are
performed using \textsc{FASTJET}~\cite{Cacciari:2011ma}.  \smallskip

\begin{figure}
\includegraphics[width=0.6\textwidth]{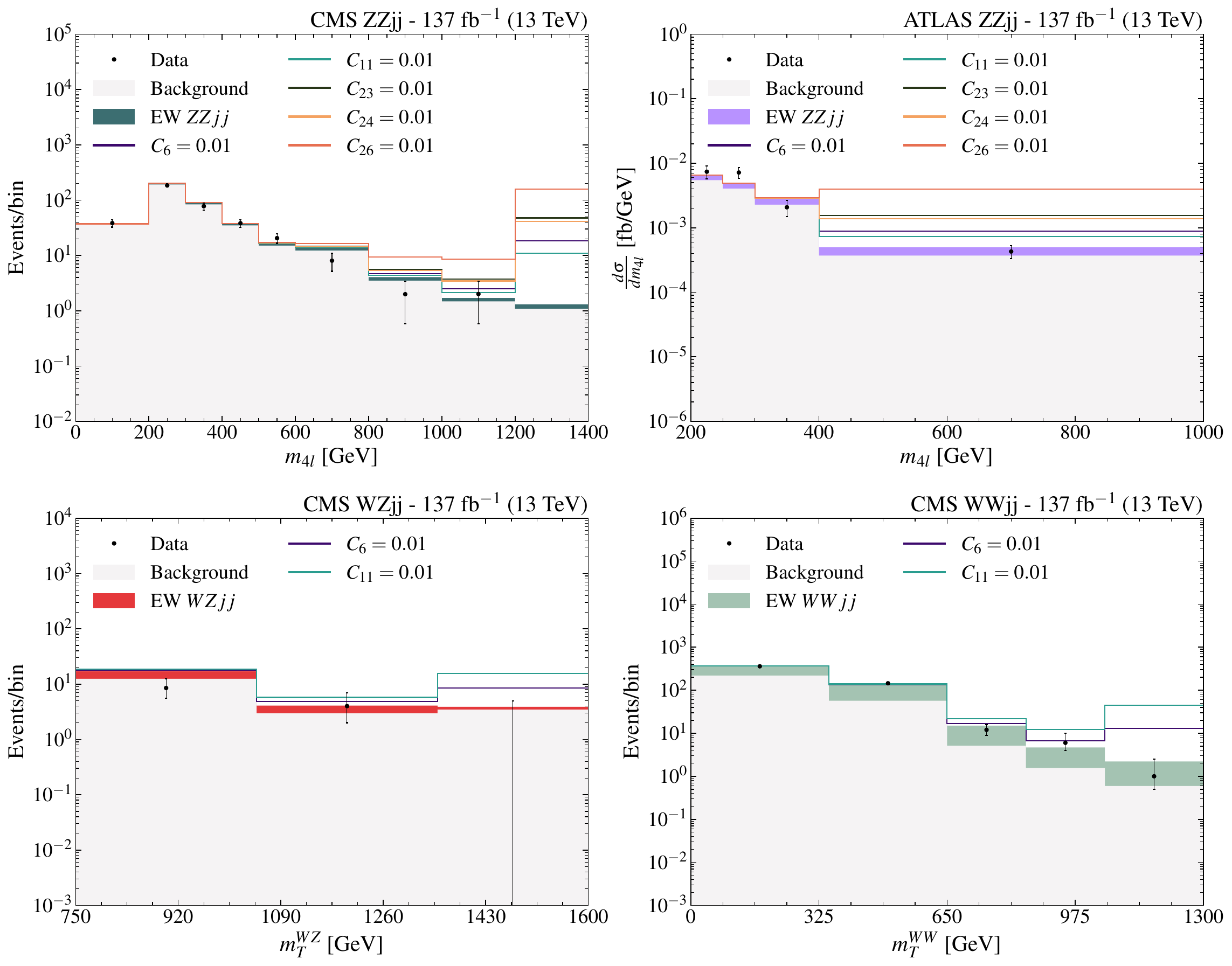}
\caption {Kinematic distributions employed in the analyses. In each
  panel we show the SM prediction, the data and the predictions for
  some illustrative values of the Wilson coefficients of the operators
  considered.}
 \label{fig:distri}
\end{figure}

For illustration, we show in Fig.~\ref{fig:distri} the kinematic
distributions used in our analyses together with the predictions for
some values of the Wilson coefficients.  As seen in this figure for
all distributions studied, the observations and SM predictions agree
with remarkable accuracy. Consequently, the data can be used to place
bounds on the new physics effects.  As expected, the effect of the new
operators is most relevant in the highest invariant mass bins. This
brings up the issue of possible violation of unitarity.  We will come
back to this point when discussing the derived bounds. \smallskip

To derive the bounds on the Wilson coefficients of the operators, we
build our test statistics $\chi^2$ function for each of the channels
following the details provided by the experimental collaborations. As
mentioned above, the experimental collaborations have performed their
searches for QGC in the framework of dimension-eight SMEFT
operators. Thus, for each channel we have tested our $\chi^2$ function
by performing first the analysis with dimension-eight SMEFT operators
to compare the sensitivity obtained with our fit and the one obtained
by the collaborations.  In this respect, it is important to notice
that both the analysis of
$W^\pm W^\pm jj\to \ell^\pm \nu \ell^{\prime\pm}\nu j j$ and
$WZ jj\to \ell^+ \ell^- \ell^{\prime}\nu j j$ events by
CMS~\cite{CMS:2020gfh} and of
$ZZjj \to \ell^+ \ell^- \ell^{\prime +}\ell^{\prime} jj$ by ATLAS
~\cite{ATLAS:2023dkz} are performed by the collaborations using
two-dimensional distributions of the invariant mass closely related to
the diboson ($M_T^{WW}$, $M_T^{WZ}$, or $m_{4\ell}$) and the dijet
invariant mass $m_{jj}$.  But there is not enough information in the
publications about the correlations between the two-dimensional
distributions to reproduce such analyses.  Therefore, we make use of
the one-dimensional distribution of the diboson-related invariant mass
and, in consequence, our bounds are consistent with those obtained by
the collaborations though slightly weaker. \smallskip

In brief:
\begin{itemize}

\item For the analysis of the CMS
  $ZZjj \to \ell^+ \ell^- \ell^{\prime +}\ell^{\prime -} jj$
  channel~\cite{CMS:2020fqz}, the number of events is large enough to
  assume gaussianity once the contents of the last four bins are
  combined.  Thus, in this case we define
  \begin{equation}
  \chi^2_{ZZ,\rm CMS} (C_6,C_{11},C_{23},C_{24},C_{26})=
  \sum_{i=1}^6 \dfrac{(N^\text{obs}_i
    - N^\text{th}_i)^2}
      {\sigma_{i}^2} 
\label{eq:chi2zzcms}
\end{equation}
where, for bin $i$, $N^\text{obs}_i$ is the observed number of events
and the expected number of events is given by
\begin{eqnarray}
  N_i^{\rm th}&=& N_i^\text{signal} +  N_i^\text{backg}
\hspace*{1cm} {\rm with} \hspace*{1cm}
  \nonumber
  N_i^\text{signal} = N_i^\text{SM} + N_i^\text{Int} + N_i^\text{BSM},
  \label{eq:nth}
\end{eqnarray}
%
%
where by $N_i^\text{SM}$, $N_i^\text{Int}$ and $N_i^\text{BSM}$ we
denote the expected number of $ZZjj$ events from the SM contribution,
the interference of the SM and HEFT ${\cal O}(p^4)$ amplitudes and the
squared amplitudes generated by the ${\cal O}(p^4)$ HEFT operators,
respectively.  $\sigma_i$ contains the statistical and uncorrelated
theoretical and systematic uncertainties added in quadrature
$\sigma_i^2=N^\text{obs}_i+(0.07\times N^\text{th})^2$.

\item For the analysis of the CMS
  $W^\pm W^\pm jj \to \ell^\pm \nu \ell^{\prime\pm}\nu j j$ process we
  use the $\chi^2$ function
  \begin{equation}
  \chi^2_{WW,\rm CMS}  (C_6,C_{11}) = 
  \min_{\Vec{\xi}} \Big\{ 2 \sum_{i=1}^5 \Big[N_i^\text{th}(\vec\xi) - N_i^\text{dat} + N_i^\text{dat} \log\Big(\dfrac{N^\text{dat}_i}{N^\text{th}_i(\vec\xi)}\Big)\Big] +
\sum_i\xi_i^2 \Big\}
\label{eq:chi2wwcms}
\end{equation}
where we introduce two pulls $\xi_1$ and $\xi_2$ to account for the
theoretical and systematic uncertainties of the signal and background
events, so
\begin{equation}
  N_i^{\rm th}(\vec\xi)= N_i^\text{signal} (1 +\sigma^{\xi_1}_i\,\xi_1) +
  N_i^\text{backg}  (1 + \sigma^{\xi_2}_i\, \xi_2) 
\end{equation}
with $\sigma^{\xi_1}_i= \sigma^{\xi_1}_i=0.07$ for $i=1,2,3$.

\item For the analysis of the CMS
  $WZ jj \to \ell^+ \ell^- \ell^{\prime}\nu j j$ events in
  Ref.~\cite{CMS:2020gfh} we focus on the last three bins from which
  we build
  \begin{equation}
  \chi^2_{WZ,\rm CMS}(C_6,C_{11},C_{23},C_{24})
  = \min_{\Vec{\xi}} \Big\{ 2 \sum_{i=1}^3 \Big[N_i^\text{th} - N_i^\text{dat} + N_i^\text{dat} \log\Big(\dfrac{N^\text{dat}_i}{N^\text{th}_i}\Big)\Big] +
\sum_i\xi_i^2\Big\}
\label{eq:chi2wzcms}
\end{equation}
with $\sigma^{\xi_1}_{i}=0.25,0.30,0.35$ and
$\sigma^{\xi_2}_{i}=0.15, 0.2, 0.25$ for $i=1,2,3$ respectively.

\item For the analysis of the ATLAS
  $ZZjj \to \ell^+ \ell^- \ell^{\prime +}\ell^{\prime -} jj$
  channel~\cite{ATLAS:2023dkz} the observable we use is the
  four-lepton invariant-mass differential cross section
  ($d\sigma/dm_{4l}$) which is a particle-level distribution, hence,
  in obtaining our predictions we need to simulate the production,
  decay, and perform the parton-shower and hadronization, but detector
  effects do not need to be simulated.  In this case we build the
  statistics
\begin{equation}
  \chi^2_{ZZ,\rm ATLAS} (C_6,C_{11},C_{23},C_{24},C_{26})
  = \sum_{i=i}^4 \dfrac{(S^\text{obs}_i- S^\text{th}_i)^2}{\sigma_{i}^2},
\label{eq:chi2zzatlas}
\end{equation}
where we read the values of $S^\text{obs}_i$ from the data points in
Fig. 4 of Ref.~\cite{ATLAS:2023dkz}. The theoretical predictions for
the differential cross section in each bin $i$ is obtained from the
generated number of events with the proper
normalization and it has the contributions
\begin{equation}
  S_i^\text{th}=S_i^\text{SM,signal} + S_i^\text{Int,signal} + S_i^\text{BSM,signal}+
  S_i^\text{backg}\;.
  \label{eq:sigatlas}
\end{equation}
The uncertainties in Eq.~\eqref{eq:chi2zzatlas}
are  $\sigma_i=(0.3,0.3,0.3,0.4)\times S^\text{obs}$  for $i=1,4$.
\end{itemize}

Finally,  we define the statistics for the global analysis
\begin{equation}
  \chi^2_{\rm GLOBAL}(C_6,C_{11},C_{23},C_{24},C_{26})=
  \chi^2_{ZZ,\rm CMS}+\chi^2_{WW,\rm CMS}+\chi^2_{WZ,\rm CMS}+\chi^2_{ZZ,\rm ATLAS}\;.
\end{equation}

\section{Results and Discussion}
\label{sec:res}

\begin{figure}
\includegraphics[width=0.6\textwidth]{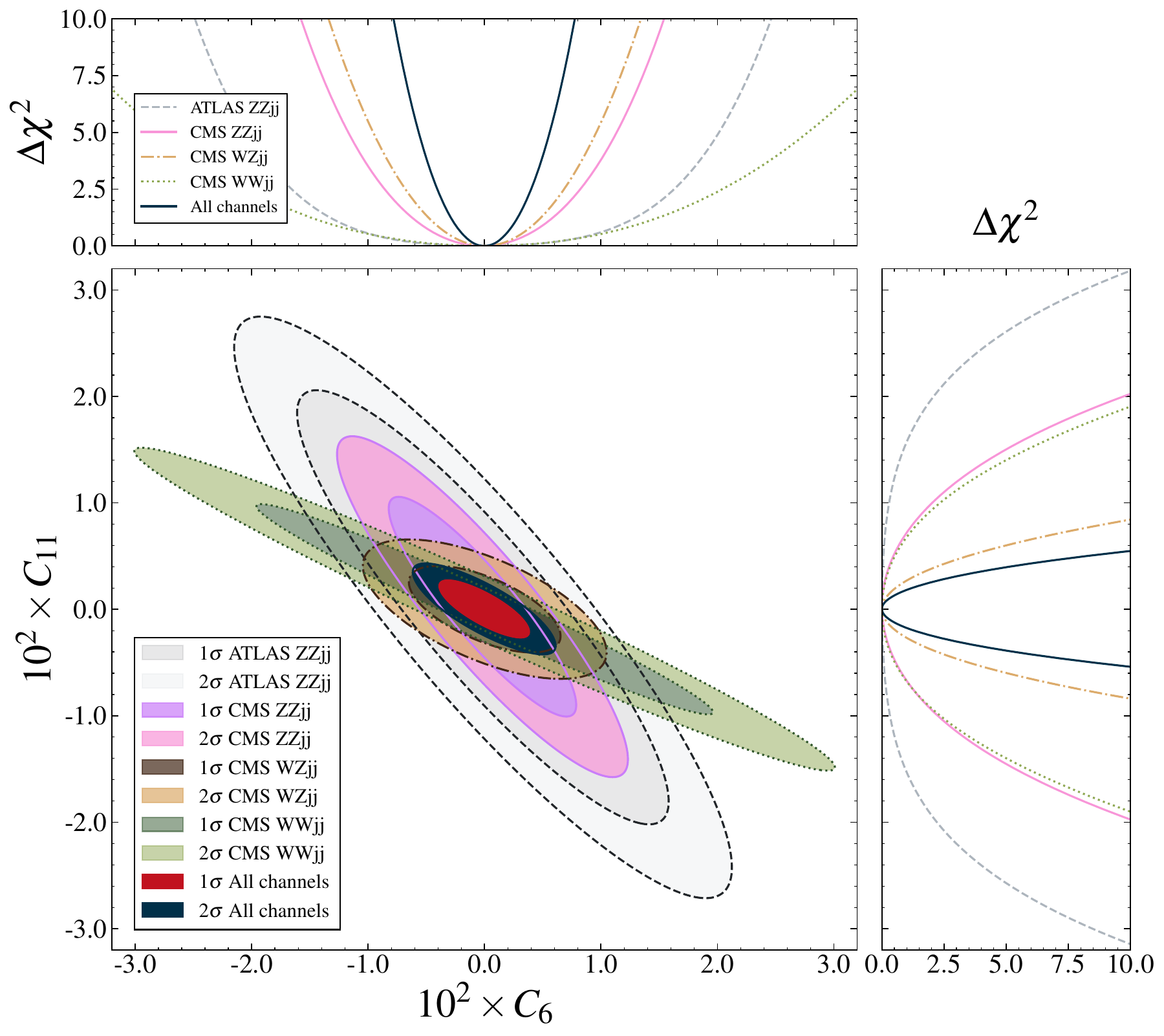}
\caption {One- and two- dimensional projections of $\Delta {\chi}^2$
  for the Wilson coefficients of the two $SU(2)_C$ conserving
  operators. We present the analyses of the different channels and
  their global combination as indicated in each panel after
  marginalizing over the undisplayed parameter.}
 \label{fig:trig2}
\end{figure}

We perform first an analysis of the data described in the previous
section including the effect of the operators which conserve the
custodial $SU(2)_C$, {\em i.e.}  ${\cal P}_6$ and ${\cal P}_{11}$.  We
plot in Fig.~\ref{fig:trig2} the 68\% and 95\% CL two-dimensional
allowed regions for their Wilson coefficients, and the corresponding
one-dimensional projections of the marginalized $\Delta\chi^2$ of the
different channels studied and their combination. From the figure we
see how the inclusion of channels involving different gauge boson
pairs is important to break the partial degeneracies between the
effect of both operators in each individual channel.  From the
one-dimensional projections we read the corresponding allowed ranges
which at 95\% CL are:
\begin{eqnarray}
  -0.0050\leq C_6\leq  0.0049\;,\\
  -0.0034\leq C_{11}\leq  0.0035\;.
\label{eq:su2cranges}  
\end{eqnarray}  

\begin{figure}
\includegraphics[width=0.9\textwidth]{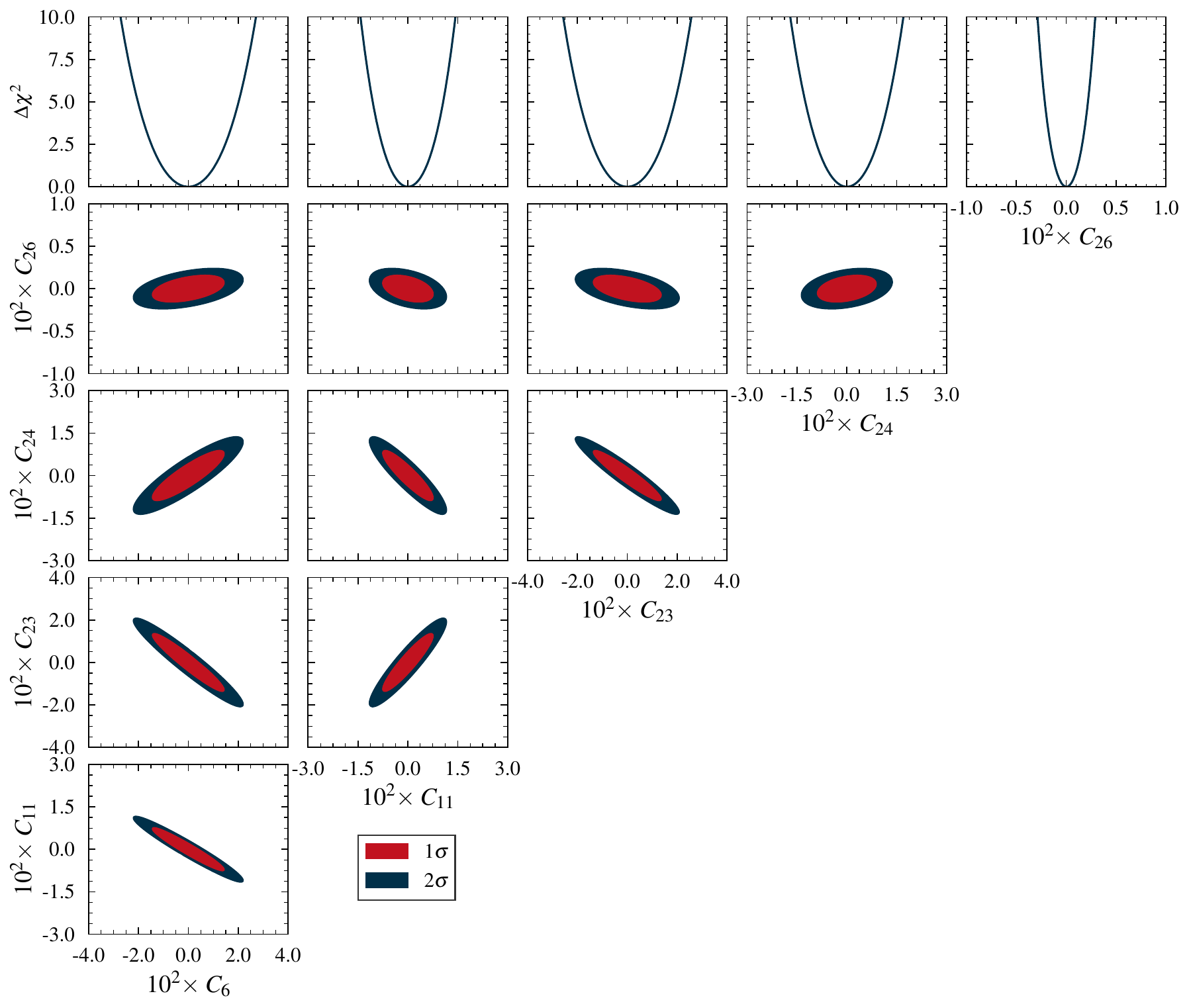}
\caption {One- and two- dimensional projections of
  $\Delta {\chi}^2_{\rm GLOBAL}$ for the Wilson coefficients of the
  five operators as indicated in each panel after marginalizing over
  the undisplayed parameters.}
 \label{fig:triglob}
\end{figure}

We then perform the analysis involving the effect of the five
operators.  In this most general case, to obtain closed bounds in the
five-dimensional parameter space, one needs to combine the data of all
channels in order to break the exact degeneracies existing in some of
the individual channels.  The results of the analysis are shown in
Fig.~\ref{fig:triglob} where we present the one- and two-dimensional
marginalized 68\% and 95\% CL allowed regions for the five Wilson
coefficients. The corresponding 95\% CL allowed ranges are listed in
the right column in Table~\ref{tab:globranges}.  As seen in the
figure, even with the combination of the four channels, there remain
large correlations or anti-correlations between $C_6$, $C_{11}$,
$C_{23}$, and $C_{24}$.  The weakest correlations occur for the
$C_{26}$ coefficient.  As a consequence, the bounds on the custodial
conserving coefficients $C_6$ and $C_{11}$ worsens by a factor
${\cal O}(4-5)$ when including the the effects of the $SU(2)_C$
violating operators in the analysis. \smallskip

\begin{table}[h]
        \centering
        \begin{tabular}{c|c|c}
          Coefficient & Individual & Marginalized\\ \hline
             $C_6$    & [-0.003, 0.003]   & [-0.018, 0.018] \\
             $C_{11}$  & [-0.002, 0.002]   & [-0.009, 0.009]   \\
             $C_{23}$  & [-0.0024, 0.0025] & [-0.017, 0.017]\\
             $C_{24}$  & [-0.0023, 0.0024] & [-0.011, 0.011]   \\
             $C_{26}$  & [-0.0013, 0.0013] & [-0.0019, 0.0020]   
        \end{tabular}
        \caption{95\% CL intervals for the Wilson coefficients in the
          analyses.  The central column are the allowed confidence
          intervals obtained by setting all other coefficients to
          zero, while the right column are the results when
          marginalizing all other four parameters in the analysis.}
        \label{tab:globranges}
    \end{table}

    For the sake of comparison, we have also performed the global
    analysis including only one operator at a time. The results are
    listed in the central column in
    Table~\ref{tab:globranges}. Comparing with the marginalized bounds
    they range from a factor 1.5 stronger for the least correlated
    coefficient, $C_{26}$, to a factor 10 tighter for
    $C_{23}$. \smallskip

\begin{figure}[h!]
\includegraphics[width=\textwidth]{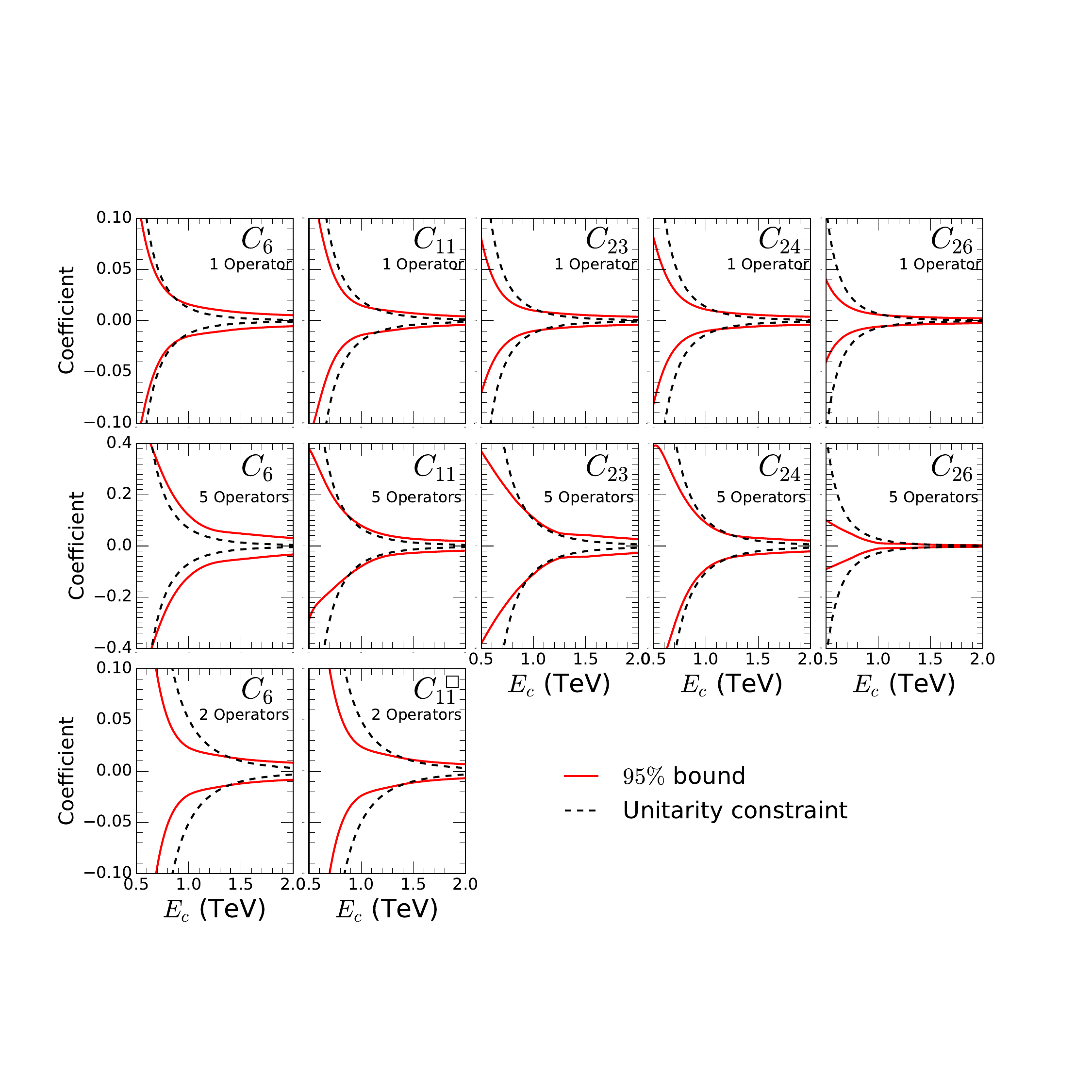}
\caption {95\% CL allowed range for the 5 Wilson coefficients from our
  analyses (full red line) as a function of the cut-off energy.  The
  upper panels correspond to the analysis in which only one
  non-vanishing operator is included in the analysis while in the
  central panels all five operators are included so the ranges shown
  in each panel are obtained after marginalization with respect to the
  other four coefficients. The lower panels depict the $SU(2)_C$
  symmetric case.  The dashed lines are the corresponding unitarity
  constraints in Eqs.\eqref{eq:unitcons} and \eqref{eq:unitcons2}.}
 \label{fig:unitarity}
\end{figure}

All the results presented so far have been obtained including the
contribution of the new operators without any constraint on the
kinematic range of the analyzed distributions.  This raises the issue
of possible violation of unitarity. In Ref.~\cite{Almeida:2020ylr} a
dedicated study of partial-wave unitarity constraints on genuine QGC
is presented for HEFT and SMEFT. The derived unitarity bounds read
\begin{equation}
  C_i\leq\{1.25,1.96,1.19.,1.35.0.69 \}
  \times 10^{-2}\times  \left(\frac{\rm TeV}{\hat s}\right)^2\, , \;\;\;
  \left[\leq\{6.9,6.9,10.,10.,2.7 \}
  \times 10^{-2}\times  \left(\frac{\rm TeV}{\hat s}\right)^2 
  \right]\,,
  \label{eq:unitcons}
\end{equation}
for $i=\{6,11,23,24,26\}$ when considering one non-vanishing operator
at a time [all five operators simultaneously] and where
$\hat s=m_{VV'}^2$ is the square of the center-of-mass (COM) energy of
the $2\rightarrow 2$ gauge-boson process. Also, with the expressions
given in Ref.~\cite{Almeida:2020ylr}, one can derive that in the
$SU(2)_C$ symmetric scenario, the unitarity bounds are
\begin{equation}
C_i\leq\{5.1,6.0 \}
  \times 10^{-2}\times  \left(\frac{\rm TeV}{\hat s}\right)^2\, ,
 \label{eq:unitcons2}
\end{equation}
for $i=\{6,11\}$. Therefore, from Table~\ref{tab:globranges} we read
that in the analysis with one operator different from zero at a time,
unitarity can be violated for the extreme values of the allowed ranges
for $\sqrt{\hat s} \geq \{1.4,1.7,1.5,1.6,1.6\}\,{\rm TeV}$ for
$i=\{6,11,23,24,26\}$ and for $\sqrt{\hat s} \geq 1.4 \,{\rm TeV}$ for
the analysis including the effect of all five operators.  In the
  $SU(2)_C$ symmetric case, the limits in Eqs.~\eqref{eq:su2cranges}
  imply that partial-wave unitarity is violated for
  $\sqrt{\hat s} \geq 1.8$ TeV. \smallskip

  Conservative bounds, which ensure unitarity conservation, can be
  obtained by repeating the analysis including the contribution of the
  anomalous operators to the observables only up to a hard kinematic
  cut-off $m_{VV'}\leq E_c$~\cite{Barger:1990py, Racco:2015dxa} and by
  studying the dependence of the derived bounds on $E_c$.  Then, the
  allowed range of coefficients is obtained for the maximum value of
  $E_c$ for which the unitarity constraint is saturated for the
  extreme values of the 95\% CL allowed range.  We plot in
  Fig.~\ref{fig:unitarity} the 95\% CL allowed range of the five
  coefficients as a function of $E_c$ compared to the unitarity bound
  for the cases with one operator is non-vanishing (upper panels),
  with all operators included (central panels) and with the inclusion
  of the $SU(2)_C$ conserving operators only (lower
  panels). \smallskip

One must notice that the unitarity constraints
Eqs.~\eqref{eq:unitcons} and ~\eqref{eq:unitcons2} do not hold a
statistical significance and therefore with this procedure one is
combining the statistically allowed ranges obtained by the analysis of
the experimental data with certain CL, with a unitarity cut-off.  So,
the values obtained with this procedure can be taken mostly as an
illustration of the loss of sensitivity when enforcing unitarity with
this method.  As seen in Fig.~\ref{fig:unitarity}, the bounds when
considering one operator at a time degrade by a factor 3-10 and by a
factor ${\cal O}(10)$ when considering all operators at a time.  In
the $SU(2)_C$ conserving scenario, the allowed ranges od $C_6$ and
$C_{11}$ become a factor $\sim 3$ and $\sim 5$ broader
respectively.\smallskip

In brief, we have obtained the bounds on genuine anomalous QGC
generated at the lowest order in the HEFT using the presently
available ATLAS and CMS experimental data on VBF production of
gauge-boson pairs.  We have considered three different scenarios
varying in the number of operators involved in the analysis.  We find
that without imposing any unitarity restriction on the anomalous cross
sections, the constraints on the Wilson coefficients are of the order
of ${\cal O}(0.003)$ TeV$^{-4}$ for scenarios in which only one
operator contributes at the time. In the $SU(2)_C$ symmetric case
[all five operators simultaneously], the limits relax to
${\cal O}(0.005)$ [${\cal O}(0.02)$] TeV$^{-4}$. Next, we restudied
the problem using a hard cut-off to guarantee that there is no
unitarity violation and obtained the most stringent constraints
without unitarity violation. Our results show that the limits on
anomalous QCG are degraded by a factor ${\cal O}(3-13)$ when we
enforce the anomalous amplitudes to respect unitarity, as expected.
The same degradation must also occur in the present limits obtained by
the experimental collaborations in the SMEFT scenario.\smallskip

\acknowledgments

We thank J. Pinheiro for his generous technical help.  OJPE is
partially supported by CNPq grant number 305762/2019-2 and FAPESP
grant 2019/04837-9.  M.M. is supported by FAPESP grant 2022/11293-8.
This project is funded by USA-NSF grant PHY-1915093.  It has also
received support from the European Union's Horizon 2020 research and
innovation program under the Marie Sk\l odowska-Curie grant agreement
No 860881-HIDDeN, and Horizon Europe research and innovation programme
under the Marie Sk\l odowska-Curie Staff Exchange grant agreement No
101086085 -- ASYMMETRY''.  It also receives support from grants
PID2019-105614GB-C21, PID2019-105614GB-C21, PID2022-136224NB-C21, and
``Unit of Excellence Maria de Maeztu 2020-2023'' award to the ICC-UB
CEX2019-000918-M, funded by MCIN/AEI/10.13039/501100011033, and from
grant 2021-SGR-249 (Generalitat de Catalunya).

\

\bibliography{references}

\end{document}